# Statistical test of Duane-Hunt's law and its comparison with an alternative law

Milan Perkovac*

*I. Technical School TESLA, Klaiceva 7, 10 000 Zagreb, Croatia*
(Dated: 26 October 2010)

**Abstract:** Using Pearson correlation coefficient a statistical analysis of Duane-Hunt and Kulenkampff's measurement results was performed. This analysis reveals that empirically based Duane-Hunt's law is not entirely consistent with the measurement data. The author has theoretically found the action of electromagnetic oscillators, which corresponds to Planck's constant, and also has found an alternative law based on the classical theory. Using the same statistical method, this alternative law is likewise tested, and it is proved that the alternative law is completely in accordance with the measurements. The alternative law gives a relativistic expression for the energy of electromagnetic wave emitted or absorbed by atoms and proves that the empirically derived Planck-Einstein's expression is only valid for relatively low frequencies. Wave equation, which is similar to the Schrödinger equation, and wavelength of the standing electromagnetic wave are also established by the author's analysis. For a relatively low energy this wavelength becomes equal to the de Broglie wavelength. Without any quantum conditions, the author made a formula similar to the Rydberg's formula, which can be applied to the all known atoms, neutrons and some hyperons.

**Résumé:** Utilisant le coefficient de corrélation de Pearson, l'auteur a fait une analyse statistique des résultats des mesures de Duane-Hunt and Kulenkampff. Cette analyse révèle qu'empiriquement la loi de Duane-Hunt n'est pas totalement conforme aux données mesurées. L'auteur a théoritiquement trouvé l'action des oscillateurs électromagnétiques, correspondant à la constante de Planck, et a aussi découvert une loi alternative basée sur la théorie classique. Utilisant la même méthode statistique, cette loi alternative est ainsi testée, et il est prouvé que cette loi alternative correspond entièrement avec les mesures. La loi alternative donne une expression relative de l'énergie des ondes électromagnétiques émises ou absorbées par les atomes et prouve que l'expression empiriquement dérivée de Planck-Einstein est seulement valide pour des fréquences relativement faibles. L'équation d'onde, qui est similaire à l'équation de Schrödinger, et la longueur d'onde de l'onde électromagnétique fixe sont aussi établies par l'analyse de l'auteur. Pour une énergie relativement faible, cette longueur d'onde devient égale à la longueur d'onde de Broglie. Sans conditions quantum, l'auteur fait une formule similaire à la formule de Rydberg, qui peut être appliquée à tous les atomes et neutrons connus, ainsi qu'à quelques hypérons.

## I. INTRODUCTION

This article is about testing Duane-Hunt's law, which was derived empirically, as well as its comparison with the alternative law, which is derived theoretically. The concept of this study is to compare Duane-Hunt's law with the measurement data, and then to compare the same measurement data with the alternative law. One of these two laws will better match the measured results. The law that better agrees with the measurements will be declared a successful law.

An important step in the realization of this article was to collect reliable experimental data. This was successfully done using the original data Duane-Hunt and Kulenkampff measured. William Duane had foreshadowed the discovery of "Duane-Hunt law" at the Philadelphia meeting of American Physical Society in December 1914, [1]. At the Washington meeting of this Society in April 1915, Duane and Hunt announced the celebrated "Duane-Hunt law". This law states that there is a sharp upper limit to the x-ray frequencies emitted from a target stimulated by impact of electrons. This frequency is determined by the equation $h\nu = eV$, independently of the material of the target; where $\nu$ is the maximum frequency of emitted radiation, $e$ is the charge of the electron, $h$ is Planck's constant, and $V$ is the total difference of potential through which the exciting electrons have fallen. Duane's physical intuition convinced him that his law could be used as a more accurate method of determining Planck's $h$, or more strictly $h/e$, than any previously used. In February 1936 R. T. Birge, well known to the physicists from 1929 for his work in establishing the best value of the physical constants, analyzing the overall situation of the fundamental physical constants, stated that of the six known methods of determining $h/e$, by far the most accurate is the method of determining the Duane-Hunt limit of the continuous spectrum.

*milan@drives.hr

However, in March 2000, I found the alternative law, [2]. This paper aims to show that the alternative law better matches with the measurements than Duane-Hunt's law. The difference between these two laws is not large at low voltages, but increases at high voltages. The comparison between Duane-Hunt's and my law is made in a wide range of voltages, between 7 000 and 32 250 V. My theoretically derived alternative law proved to be more accurate than Duane-Hunt's law. Comparison of the results is presented further in this article. Now I present my alternative law. This law is displayed in Physics Essays **15**, 41 (2002); **16**, 162 (2003), Ref. 2, and I will present it here in the short review form. The contribution of this article with respect to the previous is that the electromagnetic energy in the interior of atoms, mentioned in Ref. 2, treated here as the energy of a standing electromagnetic wave.

## II. THEORETICAL TREATMENT

The usual solution of Maxwell's equations in a limited space is a standing electromagnetic wave, [3]. This limited space can also be any atom. The existence of electromagnetic waves in the atom results in an interaction between waves and electrons. This interaction lead to quantization of the atom, [4], but quantization is beyond our interest in this article.

At the moment when the electrical energy of standing wave is on the maximum, the magnetic energy of this wave is zero, and vice versa, at the moment when the magnetic energy of standing wave is on the maximum, the electric energy of this wave is zero. Furthermore, the standing wave does not transmit energy through space, because the average active power is zero. This means that energy of standing wave oscillates with a circular frequency $\omega$ at the same location. We conclude that the energy of standing wave behaves in the same way as the energy in the $LC$ oscillatory circuit. We come to the same conclusion through the wave equation of electromagnetic fields and Lecher's transmission lines. Namely, the equations of transmission line also give the above-mentioned standing wave, [5], which acts as the energy in the $LC$ circuit. Therefore transmission line may represent a model for standing wave in the atom. According to Ref. 6, natural angular frequency of short transmission line,

$$\omega = 1/\sqrt{LC}, \qquad (1)$$

is the same as the natural angular frequency of $LC$ circuit, [7]. Therefore, transmission lines, or $LC$ circuits, can be used as a model for the describing the electromagnetic energy of standing waves in the atom. For this model to be real, it is necessary that parameters (e.g., $L$, $C$, $\delta$, $\rho$), and variables (e.g., charge $\hat{Q}_C$, or current $\hat{I}_L = \omega \hat{Q}_C$) of transmission line, i.e., $LC$ circuit, are linked to the parameters (e.g., $q$, $Q$, $m$, $\varepsilon$, $\mu$) and the variables (e.g., $r_a$, $v$, $E$, $K$, $U$, $E_{em}$) of atoms (Note: the aforementioned symbols will be explained later). Therefore, we must inspect closely the detailed Lecher's transmission line. A transmission line consists of two parallel perfect conductors with constant cross sections (twin-lead). Current flows down one conductor and returns via the other. Inductance per unit length $L'$ and the capacitance per unit length $C'$ are, [8],

$$L' = \mu \frac{\ln(\delta/\rho)+1/4}{\pi} \Big|_{\delta \gg \rho} \approx \mu \frac{\ln(\delta/\rho)}{\pi}, \qquad (2)$$

$$C' = \varepsilon \frac{\pi}{\ln\left\{\delta/(2\rho)+\sqrt{[\delta/(2\rho)]^2-1}\right\}} \Big|_{\delta \gg \rho} \approx \varepsilon \frac{\pi}{\ln(\delta/\rho)}, \qquad (3)$$

where $\mu$ and $\varepsilon$ are the magnetic permeability and the dielectric constant of the medium surrounding the conductors ($\mu = \mu_r \mu_0$, $\mu_0 = 4\pi \times 10^{-7}$ H/m, $\varepsilon = \varepsilon_r \varepsilon_0$, $\varepsilon_0 = 8.854... \times 10^{-12}$ F/m), $\rho$ is the radius of each conductor and $\delta$ is the distance between the conductors (see Lecher line, Fig. 1). The product of $L'$ and $C'$, when the distance $\delta$ is much larger than the radius $\rho$, is, [9]:

$$L'C' = \mu\varepsilon \frac{\ln(\delta/\rho)+1/4}{\ln[(\delta/\rho)/2+\sqrt{(\delta/\rho)^2/4-1}]} \Big|_{\delta \gg \rho} \approx \mu\varepsilon. \qquad (4)$$

Using (2) and (3), the characteristic impedance of $LC$ circuit, as well as the characteristic impedance of the Lecher's line reads:

$$\begin{aligned} Z_{LC} &= \sqrt{L/C} = \sqrt{L'/C'} \\ &= \frac{\sqrt{[\ln(\delta/\rho)+1/4]\ln[(\delta/\rho)/2+\sqrt{(\delta/\rho)^2/4-1}]}}{\pi}\sqrt{\mu/\varepsilon} \quad (5) \\ &= \frac{\sigma(\delta/\rho)}{\pi}\sqrt{\mu/\varepsilon}, \end{aligned}$$

where $\sigma$ from a mathematical point of view is a function of $\delta/\rho$,

$$\sigma = \sqrt{[\ln(\delta/\rho)+1/4]\ln[(\delta/\rho)/2+\sqrt{(\delta/\rho)^2/4-1}]}. \quad (6)$$

From the physical point of view $\sigma$ is a coefficient dependent on the parameters $\delta/\rho$ of the structure of transmission lines, so we will call it *structural coefficient* (see Fig.1).

Natural frequency of electromagnetic oscillator is determined by (1), i.e.,

$$\nu = 1/2\pi\sqrt{LC}. \qquad (7)$$

This means that the oscillatory period of a $LC$ oscillator is equal to

$$T = 1/\nu = 2\pi\sqrt{LC}. \qquad (8)$$

Electromagnetic energy of an $LC$ circuit, $E_{em}$, theoretically can be presented as the electrical energy stored in a capacitor that has a capacitance equal to $C$ and a maximum charge equal to $\hat{Q}_C$ (*or magnetic energy stored in a coil that has inductance equal to $L$ and maximum current equal to $\hat{I}_L$*):

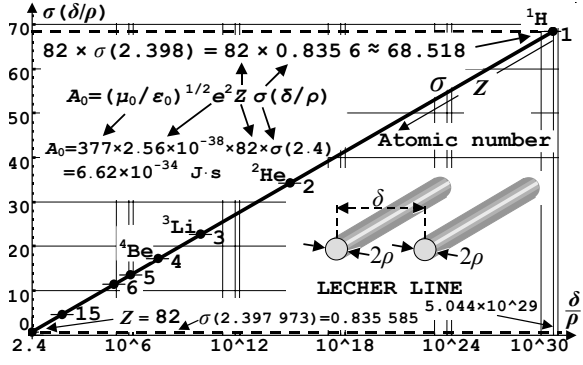

FIG. 1. Structural coefficient of twin-lead transmission line, $\sigma(\delta/\rho) = \sqrt{[\ln(\delta/\rho)+1/4]\ln[(\delta/\rho)/2+\sqrt{(\delta/\rho)^2/4-1}]}$ (*Lecher line*), versus ratio $\delta/\rho$ (*logarithmic-linear scale*).

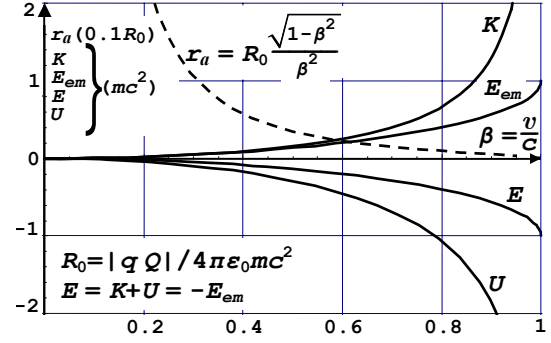

FIG. 2. Radius of the atom $r_a$, kinetic energy $K$, electromagnetic energy $E_{em}$, total mechanical energy $E$, and potential energy $U$ of the electron in an atom, versus $\beta = v/c$, ($0 \le \beta \le 1$).

$$E_{em} = \frac{1}{2}L\hat{I}_L^2 = \frac{1}{2}\frac{\hat{Q}_C^2}{C} = \frac{1}{2}\frac{\pi}{\pi}\frac{\hat{Q}_C^2}{\sqrt{C}\sqrt{C}}\frac{\sqrt{L}}{\sqrt{L}}$$
$$= \pi\sqrt{\frac{L}{C}}\hat{Q}_C^2\frac{1}{2\pi\sqrt{LC}} = \pi Z_{LC}\hat{Q}_C^2\nu = A\nu, \quad (9)$$

where

$$A = \pi Z_{LC}\hat{Q}_C^2 \quad (10)$$

is *action of electromagnetic oscillators*, defined as the product of energy, $E_{em}$, and period, $T$, of the *LC* oscillator:

$$A = E_{em}T = E_{em}/\nu. \quad (11)$$

From Newton's second law, $F=ma$, we substitte $a=v^2/r_a$ and Coulomb's law for $F$, and also from reletivistic mass we obtain:

$$\frac{m\,v^2}{r_a\sqrt{1-\beta^2}} = \frac{|qQ|}{4\pi\varepsilon r_a^2}, \quad (12)$$

or

$$r_a = \frac{|qQ|}{4\pi\varepsilon mc^2}\frac{\sqrt{1-\beta^2}}{\beta^2}, \quad (13)$$

where $r_a$ is the radius of the atom, $q$ is the charge of the electron ($q = -e$, $e = 1.602\,177...\times 10^{-19}$ C), $Q$ is the core charge, $m = 9.109...\times 10^{-31}$ kg is the electron rest mass, $c = 299\,792\,458$ m/s is the speed of light in vacuum, $\beta = v/c$, where $v$ is the electron velocity.

The kinetic energy of electrons in the atom is

$$K = \frac{mc^2}{\sqrt{1-\beta^2}} - mc^2, \quad (14)$$

and, using (13), noting that an electron is of opposite charge of the nucleus, the potential energy of electrons is

$$U = \frac{qQ}{4\pi\varepsilon r_a} = -\frac{mc^2}{\sqrt{1-\beta^2}}\beta^2. \quad (15)$$

Thus, the electromagnetic energy, $E_{em}$, (see Fig. 2) is equal to the total mehanical energy $E$ with negative sign,

$$E_{em} = -E = -(K+U) = mc^2\left(1-\sqrt{1-\beta^2}\right). \quad (16)$$

When the electromagnetic wave leaves the atom it carries out the energy $E_{em}$, and vice versa, when it enters the atom, it brings in the same energy. This means that due to the reduction of energy $E_{em}$, the atom should be compensated with equally large energy $eV$:

$$E_{em} = -E = eV. \quad (17)$$

From the expressions (16) and (17) we obtain

$$\sqrt{1-\beta^2} = 1 - eV/mc^2, \quad (18)$$

i.e.,

$$\beta^2 = (2eV/mc^2)(1 - eV/2mc^2), \quad (19)$$

or

$$\beta = \sqrt{2eV/mc^2}\sqrt{1 - eV/2mc^2}. \quad (20)$$

Using (13), (18), and (19), we obtain

$$r_a = \frac{|qQ|}{8\pi\varepsilon eV} \frac{1 - eV/mc^2}{1 - eV/2mc^2}, \qquad (21)$$

while from (21) we obtain $eV$ using (9) and (17):

$$eV = \frac{|qQ|}{8\pi\varepsilon r_a} \frac{1 - eV/mc^2}{1 - eV/2mc^2} = \frac{1}{2}\frac{\hat{Q}_C^2}{C}. \qquad (22)$$

From a single equation, (22), we need to determine two unknown sizes, i.e., parameter $C$ and variable $\hat{Q}_C$. Therefore, we seek solutions with the help of Diophantine equations. If we divide (22) with $|qQ|(1-eV/mc^2)/[2C(1-eV/2mc^2)]$, we obtain:

$$\frac{\hat{Q}_C^2}{|qQ|\frac{1-eV/mc^2}{1-eV/2mc^2}} - \frac{C}{4\pi\varepsilon r_a} = 0, \qquad (23)$$

where

$$\pm \frac{\hat{Q}_C}{\sqrt{|qQ|\frac{1-eV/mc^2}{1-eV/2mc^2}}} = x, \qquad (24)$$

$$\frac{C}{4\pi\varepsilon r_a} = y. \qquad (25)$$

Hence, we obtain Diophantine equation

$$x^2 - y = 0, \qquad (26)$$

which has many solutions. These solutions are pairs $(x, y)$ of numbers, i.e., (0,0), (1,1), (-1,1), (2,4), (-2,4), (3,9), (-3,9), ... general: $(\pm n, n^2)$, where $n = 0,1,2,3,...$. Thus, we get the following mathematically possible solutions:

$$\hat{Q}_C = n\sqrt{|qQ|\frac{1-eV/mc^2}{1-eV/2mc^2}}, \qquad (27)$$

$$C = n^2 4\pi\varepsilon r_a. \qquad (28)$$

Charge $\hat{Q}_C$ by expression (27) depends on the variables $n$, $q$, $Q$, $V$ and the constants $e$, $m$ and $c$. In a closed system, what we observe, the variables $n$, $q$ and $Q$ are fixed ($q = $ unchanging, $Q = $ unchanging, and for natural solution we choose $n = 1$). The only remaining variable is the voltage $V$. Then, we finally get variable $\hat{Q}_C$,

$$\hat{Q}_C^2 = |qQ|\frac{1-eV/mc^2}{1-eV/2mc^2}, \qquad (29)$$

and parameter $C$,

$$C = 4\pi\varepsilon r_a, [10]. \qquad (30)$$

Using (5), (7), (22) and (30) we get the following, which is similar to the expression (9):

$$eV = \frac{|qQ|}{8\pi\varepsilon r_a}\frac{1-eV/mc^2}{1-eV/2mc^2} = \frac{|qQ|}{2C}\frac{1-eV/mc^2}{1-eV/2mc^2}$$
$$= \pi Z_{LC}|qQ|\frac{1-eV/mc^2}{1-eV/2mc^2}\nu = A\nu. \qquad (31)$$

From (10) and (29) follows

$$A = \pi Z_{LC}|qQ|\frac{1-eV/mc^2}{1-eV/2mc^2}. \qquad (32)$$

If $eV = 0$ then $A = A_0$. This means that (32) can be written as

$$A = A_0 \frac{1-eV/mc^2}{1-eV/2mc^2}, \qquad (33)$$

where

$$A_0 = \pi Z_{LC}|qQ|. \qquad (34)$$

So, $A_0$ depends only on the parameters of the observed system, $Z_{LC}$, $q$ and $Q$, and is not dependent on external influences. On the other hand, there is a strong relationship between the *referent mechanical action $\underline{a}$* (*the product of energy and a period of a mechanical oscillator*) (see Ref. 2) and the *action of electromagnetic oscillator $A_0$* (*the product of energy and a period of an electromagnetic oscillator*). In Physics Essays **16**, Ref. 2, 162 (2003) I found the validity of relationship $A_0 = 2\underline{a}$, as I have shown that the referent mechanical action $\underline{a}$ is constant for all atoms

$$\{\underline{a} = \tfrac{1}{2}\sqrt{|qQ|\pi m \underline{r}/\varepsilon};$$
$$\underline{r} = Ze/4\pi\varepsilon V_i[1 + 1/(1 + eV_i/mc^2)]\},$$

where $Z$ is *atomic number*, $V_i$ is *ionization voltage of hydrogen like atoms*, i.e., 13.6 V, 54.4 V, 122.4 V, 217.7 V, 340.1 V, 489.8 V, 666.8 V, 871.1 V, 1 100 V, 1 350 V, ... for $_1^1\mathrm{H}^{1+}$, $_2^4\mathrm{He}^{2+}$, $_3^7\mathrm{Li}^{3+}$, $_4^9\mathrm{Be}^{4+}$, $_5^{11}\mathrm{B}^{5+}$, $_6^{12}\mathrm{C}^{6+}$, $_7^{14}\mathrm{N}^{7+}$, $_8^{16}\mathrm{O}^{8+}$, $_9^{19}\mathrm{F}^{9+}$, $_{10}^{20}\mathrm{Ne}^{10+}$,..., respectively, all mentioned in the literature on ionization, and $\underline{r}$ is *orbital radius of hydrogen like atoms in a referent state*}. Therefore, the physical quantity $A_0$ is a unique constant of all atoms, and in accordance with a formula which we get from the previous $\underline{a}$ and $\underline{r}$,

$$A_0 = \frac{Ze^2}{2\varepsilon} \sqrt{\frac{m}{eV_i \left(1 + \frac{1}{1 + eV_i/mc^2}\right)}} \qquad (35)$$

and from these ten ionization data, is about $6.63 \times 10^{-34}$ J·s. According to the amount that is Planck's constant ($A_0 = h = 6.626\,075... \times 10^{-34}$ J·s).

If we calculate $eV_i$ from (35), we obtain the quadratic equation,

$$(eV_i)^2 + \frac{mc^2}{A_0^2}\left(2A_0^2 - \frac{Z^2 e^4}{4\varepsilon^2 c^2}\right) eV_i - \frac{Z^2 e^4 m^2 c^2}{4\varepsilon^2 A_0^2} = 0,$$

which has two solutions, i.e., (+) and (−) solution:

$$(eV_i)_{1,2} = \frac{m}{2}\left(\frac{Ze^2}{2A_0\varepsilon}\right)^2 - mc^2$$
$$\pm \sqrt{\left[\frac{m}{2}\left(\frac{Ze^2}{2A_0\varepsilon}\right)^2\right]^2 + (mc^2)^2}.$$

Physical meaning has a solution with positive sign:

$$eV_i = \frac{m}{2}\left(\frac{Ze^2}{2A_0\varepsilon}\right)^2 - mc^2$$
$$+ \sqrt{\left[\frac{m}{2}\left(\frac{Ze^2}{2A_0\varepsilon}\right)^2\right]^2 + (mc^2)^2}, \qquad (36)$$

which, taking into account the actual values of physical quantities, gives approximately:

$$eV_i \approx \frac{m}{2}\left(\frac{Ze^2}{2A_0\varepsilon}\right)^2$$
$$= \frac{me^4}{8A_0^2\varepsilon^2} Z^2 = 13.606\, Z^2 \,[\text{eV}]. \qquad (37)$$

From (34), using (5), we get: $A_0 = \sigma\sqrt{\mu/\varepsilon}\,|qQ|$. In the atom $\sqrt{\mu/\varepsilon} = \sqrt{\mu_0/\varepsilon_0}$, so

$$A_0 = \sigma\sqrt{\mu_0/\varepsilon_0}\,|qQ|. \qquad (38)$$

From (38) we determine:

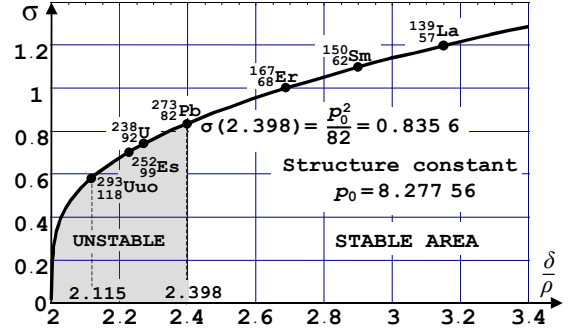

FIG. 3. Structural coefficient $\sigma = \sigma(\delta/\rho)$, versus ratio $\delta/\rho$ (stable and unstable areas).

$$\sigma = \frac{A_0}{\sqrt{\mu_0/\varepsilon_0}\,|qQ|} = \frac{A_0}{\sqrt{\mu_0/\varepsilon_0}\,Ze^2},$$
$$\sigma Z = \frac{A_0}{\sqrt{\mu_0/\varepsilon_0}\,e^2} = const. = p_0^2. \qquad (39)$$

From (6) and (39) we have:

$$\delta/\rho = e^{A_0/(\sqrt{\mu_0/\varepsilon_0}\,|qQ|)}$$
$$= e^{\sigma} = e^{p_0^2/Z}. \qquad (40)$$

From (40) we see that with increasing $|qQ| = Ze^2$, the ratio $\delta/\rho$ is falling. According to Lecher's line in Fig. 1, the theoretically lowest possible ratio $\delta/\rho$ is two, or only slightly greater than two, i.e., $\delta \geq 2\rho$. Atoms with a smaller ratio than this (i.e., $\delta/\rho < 2$), in accordance with the model of Lecher's line, cannot exist. There are atoms whose ratio $2 \leq \delta/\rho \leq 2.4$. As we shall see immediately, these atoms are unstable. If we review the data on the stability of all the known atoms [11], we see that the stabile atom with the highest atomic number is lead, $^{207}_{82}\text{Pb}$. This means that the atomic number corresponding to the ratio $\delta/\rho \approx 2.4$ is $Z = 82$, i.e., $Q = 82e$, (Fig. 3). From (6) and (40) we get:

$$A_0 = \sqrt{\mu_0/\varepsilon_0}\,|qQ|\sigma(\delta/\rho)$$
$$= \sqrt{\mu_0/\varepsilon_0}\,Ze^2\sigma(\delta/\rho), \qquad (41)$$

and using (39), $\sqrt{\sigma Z} = p_0$,

$$A_0 = \sqrt{\mu_0/\varepsilon_0}\,e^2 p_0^2 = \sqrt{\mu_0/\varepsilon_0}\,(ep_0)^2,$$

where $ep_0 = q_0$ we call *charge of the structure*, and $p_0$ we call *the constant of the charge of the structure*, or

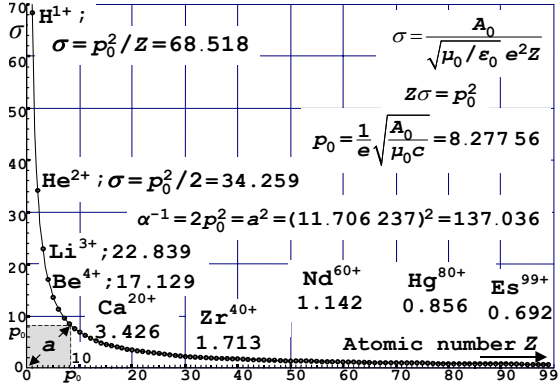

FIG. 4. Structural coefficient $\sigma = p_0^2 / Z$, versus atomic number Z (*equilateral hyperbole of atoms*).

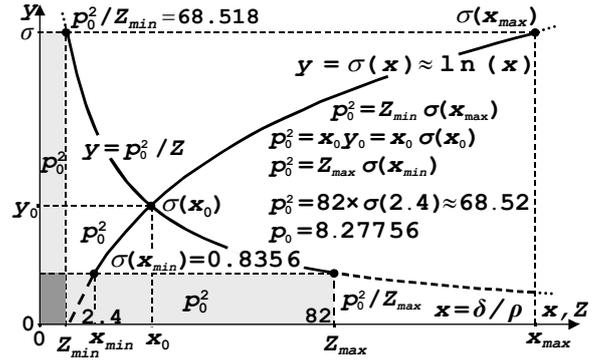

FIG. 5. Structural coefficient in the form $\sigma = \sigma(\delta / \rho)$, and structural coefficient in the form $\sigma = p_0 / Z$, versus the same time $x = \delta / \rho$ and Z.

(simply) *the structure constant*. Thus, $A_0$ is determined by the universal constants $\mu_0$, $\varepsilon_0$ and $e$, and structure constant $p_0$. Using (41), with the help of these lead data, we can now estimate $A_0$ (see Fig. 3):

$A_0 = 376.73 \text{ V} \cdot \text{A}^{-1} \cdot 82 \cdot (1.602177 \cdot 10^{-19} \text{ A} \cdot \text{s})^2 \cdot \sigma(2.398)$
$= 6.626 \times 10^{-34} \text{ V} \cdot \text{A} \cdot \text{s}^2 = 6.626 \times 10^{-34} \text{ J} \cdot \text{s}$,

where $\sigma(2.398)$, in accordance with (6), is $\sigma(2.398) = 0.8356$.

The equation (39), $\sigma Z = p_0^2$, represents the hyperbola; mathematical $xy = a^2 / 2$. Such hyperbole has no extremes. From previous equations derived: $p_0^2 = 1/2\alpha$, i.e., $a = 1/\sqrt{\alpha} = 11.706$, or $\alpha^{-1} = a^2$, where $\alpha$ is a fine-structure constant. Thus, the curve $\sigma Z = 1/2\alpha = 68.518$ represents the *equilateral hyperbola of atoms* (Fig. 4). From (39) and $c = 1/\sqrt{\mu_0 \varepsilon_0}$ we get:

$$p_0 = \frac{1}{e}\sqrt{\frac{A_0}{\mu_0 c}} = 8.27756. \qquad (42)$$

Structure constant $p_0$ is defined independently of the Planck constant and fine-structure constant. Planck's constant in (42) ($A_0 = h$) is used only for accurately calculating the structure constant $p_0$. Introducing dimensionless structure constants, $p_0 = 8.27756$, changing the structure of several other universal physical constants (see Table I).

The relationships in terms of structure constant $p_0$ are seen once again on Fig. 5, which was created by consolidating the Fig. 1 and Fig. 4.

From (31) follows $eV = A\nu$, and using (33) we get:

$$\nu = \frac{eV}{A_0} \frac{1 - eV/2mc^2}{1 - eV/mc^2}. \qquad (43)$$

Expression (43) represents the alternative law. This alternative law includes a relativistic theory. For $eV \ll mc^2$, and $A_0 = h$, this alternative law converges into Duane-Hunt's law,

$$\nu = eV/h. \qquad (44)$$

TABLE I. Some of the universal physical constants that depend on a structure constant $p_0$ (proton mass = $m_p$).

| Quantity | Symbol | Form and value |
|---|---|---|
| Structure constant | $p_0$ | 8.27756 |
| Charge of the structure | $q_0$ | $ep_0 = 1.326212 \times 10^{-18}$ C |
| Fine-structure constant | $\alpha$ | $1/\alpha = 2p_0^2 = 137.035999$ |
| Planck constant | $h$ | $(\mu_0/\varepsilon_0)^{1/2}(ep_0)^2 = 6.626076 \times 10^{-34}$ J·s |
| Rydberg constant | $R$ | $m/8\mu_0 e^2 p_0^6 = 1.097373 \times 10^7$ m$^{-1}$ |
| Bohr radius | $a$ | $\mu_0 e^2 p_0^4/\pi m = 5.291773 \times 10^{-11}$ m |
| Bohr magneton | $\mu_B$ | $(\mu_0/\varepsilon_0)^{1/2} e^3 p_0^2/4\pi m$ |
|  |  | $= 9.274016 \times 10^{-24}$ A·m$^2$ |
| Nuclear magneton | $\mu_N$ | $(\mu_0/\varepsilon_0)^{1/2} e^3 p_0^2/4\pi m_p$ |
|  |  | $= 5.050786 \times 10^{-27}$ A·m$^2$ |

## III. STATISTICAL PROCESSING

Now, we move on to the statistical processing. As stated in Ref. 12, between two random variables, the correlation is a measure of the extent to which a change in one tends to correspond to a change in the other. The correlation is high or low depending on whether the relationship between the two variables is close or not. If the change in one corresponds to a change in the other in the same direction, there is a positive correlation, and there is a negative correlation if the changes are in opposite directions.

Independent random variables have zero correlation. Zero correlation also always appears when the results from the correct theory, $v_{theor}$, are subtracted from the results of correct measurement, $v_{expt}$, thus, when $Y = v_{expt} - v_{theor}$. So it is possible to verify the correctness of each theory, i.e., to quantify its agreement with the measurements. If the theory is correct, and coincides with the measurements, the correlation will be around zero.

One measure of correlation between the random variables $X$ and $Y$ is *correlation coefficient* $r$. Note that the existence of some correlation between two variables need not imply that the link between them is causal. For a sample of $n$ paired observations $(x_1, y_1)$, $(x_2, y_2)$, …, $(x_n, y_n)$, the Pearson's coefficient of linear correlation is equal to, [13]:

$$r = \frac{\sum_{i=1}^{n} x_i y_i - n\, \bar{x}\, \bar{y}}{\sqrt{\left(\sum_{i=1}^{n} x_i^2 - n\, \bar{x}^2\right)\left(\sum_{i=1}^{n} y_i^2 - n\, \bar{y}^2\right)}}. \quad (45)$$

This satisfies $-1 \leq r \leq 1$. If $X$ and $Y$ are linearly related, then $r = -1$ or $+1$. Meanings of various symbols in expression (45) are best seen in Table II, and the next ten relations, (46)-(56).

In our case, *independent variable* $X$ is a voltage ($X = V$). *Dependent variable* $Y$ is the difference between measured ($v_{expt}$) and calculated ($v_{theor}$) frequencies, i.e., $Y_1 = v_{expt} - v_{theor1}$ (for calculation of the coefficient of linear correlation $r_1$), and $Y_2 = v_{expt} - v_{theor2}$ (for calculation of the coefficient of linear correlation $r_2$).

Number of samples in the observed case is fourteen ($n = 14$). For the independent variable $X$ we have:

$$\bar{x} = \frac{1}{n}\sum_{i=1}^{n} x_i = \frac{190\,700\ \text{V}}{14} = 13\,621.43\ \text{V}, \quad (46)$$

$$\bar{x}^2 = (13\,621.43\ \text{V})^2 = 185\,543\,355.20\ \text{V}^2, \quad (47)$$

$$\sum_{i=1}^{n} x_i^2 = 350.83 \times 10^7\ \text{V}^2. \quad (48)$$

In the case of dependent variables $Y_1$, subtrahend $v_{theor1}$ in Table II is specified by Duane-Hunt law, (44), i.e., $v_{theor1} = eV/h$, in which $v_{theor1}$ is the theoretical maximum frequency of x-rays emitted from the target according to this law (Planck constant $h$ is $6.6261 \times 10^{-34}$ J·s).

TABLE II. Data for calculating Pearson linear correlation coefficients $r_1$ and $r_2$, as well as to calculate the linear regression $\hat{y} = \hat{\beta}x + \hat{\alpha}$. Experimental results in Refs.14 and 15.   $Y_1 = v_{expt} - v_{theor1}$;   $Y_2 = v_{expt} - v_{theor2}$.

| $i$ | $X$ | $X^2$ | $v_{expt}$ | $v_{theor1}$ [a] | $Y_1$ | $Y_1^2$ | $XY_1$ | $v_{theor2}$ [b] | $Y_2$ | $Y_1^2$ | $XY_2$ |
|---|---|---|---|---|---|---|---|---|---|---|---|
| | $(10^0)$ | $(10^7)$ | $(10^{18})$ | $(10^{18})$ | $(10^{16})$ | $(10^{32})$ | $(10^{20})$ | $(10^{18})$ | $(10^{16})$ | $(10^{32})$ | $(10^{20})$ |
| | (V) | (V$^2$) | (Hz) | (Hz) | (Hz) | (Hz$^2$) | (V·Hz) | (Hz) | (Hz) | (Hz$^2$) | (V·Hz) |
| 1 | 7 000 | 4.900 | 1.705[c] | 1.6923 | 1.241 | 1.540 | 0.869 | 1.704 | 0.096 3 | 0.001 | 0.006 74 |
| 2 | 7 000 | 4.900 | 1.710[c] | 1.6936 | 1.741 | 3.030 | 1.219 | 1.704 | 0.596 3 | 0.250 | 0.417 43 |
| 3 | 7 850 | 6.162 | 1.912[c] | 1.898 | 1.388 | 1.926 | 1.090 | 1.913 | −0.058 1 | 0.063 | −0.045 59 |
| 4 | 8 750 | 7.656 | 2.131[c] | 2.116 | 1.526 | 2.329 | 1.335 | 2.134 | −0.278 2 | 0.281 | −0.243 46 |
| 5 | 8 750 | 7.656 | 2.130[c] | 2.116 | 1.426 | 2.034 | 1.248 | 2.134 | −0.378 2 | 0.593 | −0.330 96 |
| 6 | 9 600 | 9.216 | 2.340[c] | 2.321 | 1.873 | 3.509 | 1.798 | 2.343 | −0.306 6 | 1.210 | −0.294 29 |
| 7 | 10 470 | 10.96 | 2.555[c] | 2.532 | 2.337 | 5.460 | 2.446 | 2.558 | −0.264 8 | 0.292 | −0.277 22 |
| 8 | 10 470 | 10.96 | 2.550[c] | 2.532 | 1.837 | 3.373 | 1.923 | 2.558 | −0.764 8 | 2.132 | −0.800 72 |
| 9 | 11 200 | 12.54 | 2.738[c] | 2.708 | 2.985 | 8.912 | 3.344 | 2.738 | 0.000 7 | 0.980 | 0.000 73 |
| 10 | 11 980 | 14.35 | 2.927[c] | 2.897 | 3.025 | 9.151 | 3.624 | 2.931 | −0.398 9 | 0.000 | −0.477 93 |
| 11 | 11 980 | 14.35 | 2.928[c] | 2.8978 | 3.125 | 9.766 | 3.744 | 2.931 | −0.298 9 | 0.792 | −0.358 13 |
| 12 | 25 000 | 62.50 | 6.169[a] | 6.045 | 12.40 | 153.8 | 31.01 | 6.199 | −3.032 4 | 9.181 | −7.580 91 |
| 13 | 28 400 | 80.66 | 7.054[a] | 6.867 | 18.69 | 349.4 | 53.08 | 7.068 | −1.386 1 | 1.904 | −3.936 62 |
| 14 | 32 250 | 104.00 | 8.081[a] | 7.798 | 28.30 | 800.8 | 91.26 | 8.059 | 2.180 1 | 4.753 | 7.030 93 |
| $\sum_{i=1}^{14}$ | 190 700 | 350.83 | | | 81.90 | 1 355.05 | 197.99 | | −4.293 6 | 17.456 | −6.829 30 |

[a]Reference 14.
[b]Reference 2.
[c]Reference 15.

$$\overline{y}_1 = \frac{1}{n}\sum_{i=1}^{n} y_{1i} = \frac{81.90 \times 10^{16} \text{ Hz}}{14} = 5.85 \times 10^{16} \text{ Hz}, \quad (49)$$

$$\overline{y}_1^2 = (5.85 \times 10^{16} \text{ Hz})^2 = 3.42 \times 10^{33} \text{ Hz}^2, \quad (50)$$

$$\sum_{i=1}^{n} y_{1i}^2 = 1355 \times 10^{32} \text{ Hz}^2, \quad (51)$$

$$\sum_{i=1}^{n} x_i y_{1i} = 197.99 \times 10^{20} \text{ V} \cdot \text{Hz}. \quad (52)$$

In the case of dependent variables $Y_2$, subtrahend $\nu_{theor2}$ in Table II is specified by alternative law, (43), i.e., $\nu_{theor2} = eV(1 - eV/2mc^2)/[A_0(1 - eV/mc^2)]$.

Here, $\nu_{theor2}$ is the theoretical maximum frequency of x-rays emitted from the target according to (43), and $A_0$, according to this paper, and Ref. 2, is also constant as is Planck's constant $h$ (i.e., $A_0 = h$).

$$\overline{y}_2 = \frac{1}{n}\sum_{i=1}^{n} y_{2i} = \frac{-4.2936 \times 10^{16} \text{ Hz}}{14}$$
$$= -0.30668 \times 10^{16} \text{ Hz}, \quad (53)$$

$$\overline{y}_2^2 = (-0.30668 \times 10^{16} \text{ Hz})^2$$
$$= 0.00940526 \times 10^{33} \text{ Hz}^2, \quad (54)$$

$$\sum_{i=1}^{n} y_{2i}^2 = 17.456 \times 10^{32} \text{ Hz}^2, \quad (55)$$

$$\sum_{i=1}^{n} x_i y_{2i} = -6.82930 \times 10^{20} \text{ V} \cdot \text{Hz} \quad (56)$$

Now all these parameters can be included in (45). Thus, we obtain, in the case of Duane-Hunt's law, correlation coefficient $r_1$:

$$r_1 = \frac{\sum_{i=1}^{n} x_i y_{1i} - n \overline{x} \overline{y}_1}{\sqrt{\left(\sum_{i=1}^{n} x_i^2 - n \overline{x}^2\right)\left(\sum_{i=1}^{n} y_{1i}^2 - n \overline{y}_1^2\right)}} = 0.97. \quad (57)$$

In the same way we obtain, in case of alternative law, correlation coefficient $r_2$:

$$r_2 = \frac{\sum_{i=1}^{n} x_i y_{2i} - n \overline{x} \overline{y}_2}{\sqrt{\left(\sum_{i=1}^{n} x_i^2 - n \overline{x}^2\right)\left(\sum_{i=1}^{n} y_{2i}^2 - n \overline{y}_2^2\right)}} = -0.08. \quad (58)$$

When we subtract the results of correct theory (*means a theory that is consistent with the measurements*) from the results of measurement, only independent random variables remain. Its correlation coefficient should be around zero. So, a big correlation coefficient of $r_1 = 0.97$ specifies that the empirically derived Duane-Hunt law, (44), does not have correct functional dependence.

Therefore, the correct functional dependence, with a correlation coefficient around zero, $r_2 = -0.08$, is determined by alternative law, i.e., Eq. (43).

Regression equation,

$$\hat{y} = \hat{\beta}x + \hat{\alpha}, \quad (59)$$

is determined using least squares and calculated, [16]:

$$\hat{\beta} = \frac{\sum_{i=1}^{n} x_i y_i - n \overline{x} \overline{y}}{\sum_{i=1}^{n} x_i^2 - n \overline{x}^2}, \quad (60)$$

$$\hat{\alpha} = \overline{y} - \hat{\beta}\overline{x}. \quad (61)$$

So in the case of Duane-Hunt's law we get,

$$\hat{\beta}_1 = \frac{\sum_{i=1}^{n} x_i y_{1i} - n \overline{x} \overline{y}_1}{\sum_{i=1}^{n} x_i^2 - n \overline{x}^2} = 9.49 \times 10^{12} \frac{\text{Hz}}{\text{V}}, \quad (62)$$

$$\hat{\alpha}_1 = \overline{y}_1 - \hat{\beta}_1 \overline{x} = -7.08 \times 10^{16} \text{ Hz}, \quad (63)$$

and in the case of the alternative law,

$$\hat{\beta}_2 = \frac{\sum_{i=1}^{n} x_i y_{2i} - n \overline{x} \overline{y}_2}{\sum_{i=1}^{n} x_i^2 - n \overline{x}^2} = -1.08 \times 10^{11} \frac{\text{Hz}}{\text{V}}, \quad (64)$$

$$\hat{\alpha}_2 = \overline{y}_2 - \hat{\beta}_2 \overline{x} = -1.60 \times 10^{15} \text{ Hz}. \quad (65)$$

Everything mentioned is illustrated in Fig. 6. This figure was created with the help of previously calculated parameters, and also with the help of *Mathematica*, Wolfram Research.

It has been proved with the help of correlation coefficients $r_1$ and $r_2$, and with the help of regression lines $\hat{y}_1$ and $\hat{y}_2$ (see Fig. 6), Duane-Hunt's and Kulenkampf's results of measurements best describes equation (43), i.e., the alternative law $\nu = eV(1 - eV/2mc^2)/[A_0(1 - eV/mc^2)]$.

According to these results, the Eq. (43) is the real term correlation of the maximum emitted frequency and applied voltage, and not Duane-Hunt law, which is represented in Eq. (44).

Increased deviation of measurement results in relation to the Duane-Hunt's law is expected at higher voltages. These deviations we checked from 7 kV to 32.25 kV. Now, it would be desirable to create and test voltages significantly higher than that.

Here, we present an alternative law, which is strictly based on theoretical grounds. This alternative law fully complies with the measurement results and successfully replaces empirically derived Duane-Hunt's law.

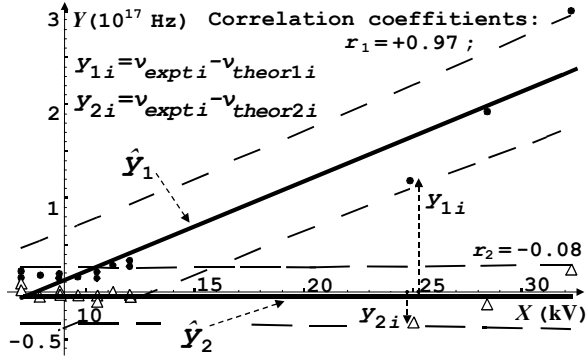

FIG. 6. Scattering diagrams of voltage ($X$), and frequency differences ($Y_1$) calculated with Duane-Hunt's law, i.e., function (44), marked ●, as well as frequency differences ($Y_2$), calculated with the alternative law, i.e., function (43), marked Δ, and also regression lines for both cases, $\hat{y}_1 = 9.49 \times 10^{12} x - 7.08 \times 10^{16}$, and $\hat{y}_2 = -1.08 \times 10^{11} x - 1.06 \times 10^{15}$, respectively (solid curves).

### IV. OTHER IMPLICATIONS OF THE ALTERNATIVE LAW

Furthermore, from (17) and (43) we get the general relativistic expression for the energy of electromagnetic waves emitted or absorbed by the atom,

$$E_{em} = A_0 \nu + mc^2 - \sqrt{(A_0 \nu)^2 + (mc^2)^2} . \quad (66)$$

If the frequency is low, then

$$E_{em} |_{\nu \to 0} = A_0 \nu . \quad (67)$$

This, in the case of $A_0 = h$, proves the validity of empirically derived Planck-Einstein's relation $E_{em} = h\nu$, but also shows that Planck-Einstein's relation is limited only to relatively low frequencies $\nu$. If this frequency is high, then

$$E_{em} |_{\nu \to \infty} = mc^2 . \quad (68)$$

From (9) and (66) follows the action of electromagnetic oscillators:

$$A = A_0 + mc^2 / \nu - \sqrt{A_0^2 + (mc^2/\nu)^2} . \quad (69)$$

We see that the action of electromagnetic oscillators, $A$, is not constant. This suggests that Planck's $h$ probably is not constant for all frequencies $\nu$.

Momentum of electromagnetic wave, $p_{em}$, in accordance with Ref. 17, is equal to the ratio of its energy

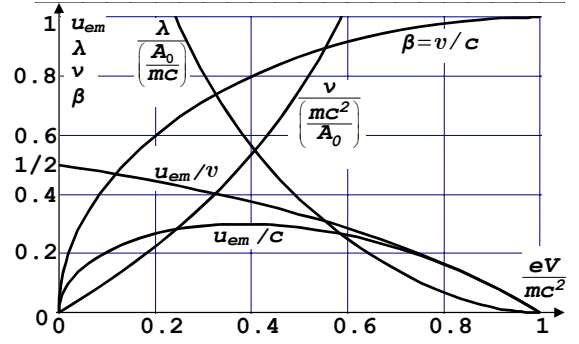

FIG. 7. Phase velocity $u_{em}$, wavelength $\lambda$ and frequency $\nu$ of standing-electromagnetic wave, also the speed of the particles $\beta = v/c$, all versus $eV/mc^2$.

($E_{em}$) and phase velocity

$$u_{em} = \lambda \nu , \quad (70)$$

i.e.,

$$p_{em} = E_{em} / u_{em} = E_{em} / \lambda \nu, \quad (71)$$

where $\lambda$ is the wavelength of the electromagnetic wave. On the other hand, in accordance with the law of conservation of momentum, the momentum of electrons is equal to the momentum of the electromagnetic wave,

$$E_{em}/\lambda \nu = m v / \sqrt{1-\beta^2} = mc\beta / \sqrt{1-\beta^2} . \quad (72)$$

From the expression (72), and using (9), (18), (20), (22), and (33), we obtain

$$\lambda = \frac{A_0}{\sqrt{2meV}} \frac{(1 - eV/mc^2)^2}{\sqrt{(1 - eV/2mc^2)^3}} . \quad (73)$$

If the voltage is low, then $eV \approx \tfrac{1}{2} m v^2$, and

$$\lambda |_{eV/mc^2 \to 0} = \frac{A_0}{m v} . \quad (74)$$

If the voltage approaches $mc^2 / e \approx 511$ kV, then

$$\lambda |_{eV/mc^2 \to 1} \to 0 \text{ m} . \quad (75)$$

Standing-wave phase velocity, according to (43), (70) and (73), is (see Fig. 7)

$$u_{em} = \lambda \nu = \sqrt{\frac{eV}{2m}} \frac{1 - eV/mc^2}{\sqrt{1 - eV/2mc^2}} . \qquad (76)$$

If the voltage is low, then

$$u_{em}|_{eV/mc^2 \to 0} = \sqrt{eV/2m}$$
$$= \sqrt{\tfrac{1}{2}mv^2/(2m)} = v/2 . \qquad (77)$$

On the other hand, Maxwell's equations also require the following relation to be satisfied

$$u_{em} = 1/\sqrt{\mu\varepsilon} , \qquad (78)$$

while the impedance $\sqrt{\mu/\varepsilon}$, because of energy reasons, should remain unchanged, i.e.,

$$\sqrt{\mu/\varepsilon} = \sqrt{\mu_0/\varepsilon_0} . \qquad (79)$$

From (76) and (78) we get

$$1/\sqrt{\mu\varepsilon} = \sqrt{\frac{eV}{2m}} \frac{1 - eV/mc^2}{\sqrt{1 - eV/2mc^2}} . \qquad (80)$$

System of equations (79) and (80) have solutions $\mu$ and $\varepsilon$ as follows:

$$\mu = \sqrt{\frac{\mu_0}{\varepsilon_0}} \sqrt{\frac{2m}{eV}} \frac{\sqrt{1 - eV/2mc^2}}{1 - eV/mc^2} = \mu_r \mu_0 , \qquad (81)$$

$$\varepsilon = \sqrt{\frac{\varepsilon_0}{\mu_0}} \sqrt{\frac{2m}{eV}} \frac{\sqrt{1 - eV/2mc^2}}{1 - eV/mc^2} = \varepsilon_r \varepsilon_0 , \qquad (82)$$

i.e.,

$$\mu_r = \varepsilon_r = \frac{1}{\sqrt{\mu_0 \varepsilon_0}} \sqrt{\frac{2m}{eV}} \frac{\sqrt{1 - eV/2mc^2}}{1 - eV/mc^2} . \qquad (83)$$

Current values of linearly polarized standing wave, [18], in the atom read:

$$\mathcal{E}_x(z,t) = \mathcal{E}_0 \sin\left(\frac{2\pi}{\lambda} z\right) \cos(\omega t) , \qquad (84)$$

$$\mathcal{H}_y(z,t) = -\frac{\mathcal{E}_0}{\sqrt{\mu/\varepsilon}} \cos\left(\frac{2\pi}{\lambda} z\right) \sin(\omega t) , \qquad (85)$$

where $\mathcal{E}_0$ is the maximum value, i.e., the amplitude of electric field strength, $\mathcal{E}_x(z,t)$, the x-component of the electric field dependent on the z-axis of the rectangular coordinate system and the time $t$, and $\mathcal{H}_y(z,t)$ is the y-component of the magnetic field dependent on the z-axis of the rectangular coordinate system and the time $t$. If we use the second derivation of the equations (84) and (85) with respect to z, we get equations:

$$\frac{\partial^2 \mathcal{E}_x(z,t)}{\partial z^2} + \left(\frac{2\pi}{\lambda}\right)^2 \mathcal{E}_x(z,t) = 0 , \qquad (86)$$

$$\frac{\partial^2 \mathcal{H}_y(z,t)}{\partial z^2} + \left(\frac{2\pi}{\lambda}\right)^2 \mathcal{H}_y(z,t) = 0 . \qquad (87)$$

After inclusion of wavelengths, (73), in the expressions (86) and (87), we get:

$$\frac{\partial^2 \mathcal{E}_x(z,t)}{\partial z^2} + \frac{8\pi^2 meV}{A_0^2} \frac{(1 - eV/2mc^2)^3}{(1 - eV/mc^2)^4} \mathcal{E}_x(z,t) = 0 , (88)$$

$$\frac{\partial^2 \mathcal{H}_y(z,t)}{\partial z^2} + \frac{8\pi^2 meV}{A_0^2} \frac{(1 - eV/2mc^2)^3}{(1 - eV/mc^2)^4} \mathcal{H}_y(z,t) = 0. (89)$$

Since $eV = -E$, we can write,

$$\frac{\partial^2 \mathcal{E}_x(z,t)}{\partial z^2} - \frac{8\pi^2 mE}{A_0^2} \frac{(1 + E/2mc^2)^3}{(1 + E/mc^2)^4} \mathcal{E}_x(z,t) = 0, \qquad (90)$$

$$\frac{\partial^2 \mathcal{H}_y(z,t)}{\partial z^2} - \frac{8\pi^2 mE}{A_0^2} \frac{(1 + E/2mc^2)^3}{(1 + E/mc^2)^4} \mathcal{H}_y(z,t) = 0, \qquad (91)$$

while in the case of low-energy ($eV = \tfrac{1}{2} mv^2 = K = E - U$),

$$\frac{\partial^2 \mathcal{E}_x(z,t)}{\partial z^2} + \frac{8\pi^2 m}{A_0^2} (E - U) \mathcal{E}_x(z,t) = 0, \qquad (92)$$

$$\frac{\partial^2 \mathcal{H}_y(z,t)}{\partial z^2} + \frac{8\pi^2 m}{A_0^2} (E - U) \mathcal{H}_y(z,t) = 0, \qquad (93)$$

which in this form resembles the Schrödinger's equation. Unlike wave function $\Psi$ in the Schrödinger's equation, $\partial^2 \Psi / \partial x^2 + 8\pi^2 m(E-U)\Psi/h^2 = 0$, [19], physical meaning of wave functions, $\mathcal{E}_x$ and $\mathcal{H}_y$, in this article is entirely clear. Meaning of all the other physical quantities in this article is also completely clear.

The solutions of differential equations (92) and (93) are the same as the known solutions Schrödinger's equation, only the meaning of the wave function is different.

In the same way as we got (86) and (87), using the second derivation with respect to time $t$, with $\omega = 2\pi\nu$, we get equations:

$$\frac{\partial^2 \mathcal{E}_x(z,t)}{\partial t^2} + \omega^2 \mathcal{E}_x(z,t) = 0, \quad (94)$$

$$\frac{\partial^2 \mathcal{H}_y(z,t)}{\partial t^2} + \omega^2 \mathcal{H}_y(z,t) = 0, \quad (95)$$

or, using (43), and $eV = -E$,

$$\frac{\partial^2 \mathcal{E}_x(z,t)}{\partial t^2} + \frac{4\pi^2 E^2}{A_0^2}\left(\frac{1 + E/2mc^2}{1 + E/mc^2}\right)^2 \mathcal{E}_x(z,t) = 0, \quad (96)$$

$$\frac{\partial^2 \mathcal{H}_y(z,t)}{\partial t^2} + \frac{4\pi^2 E^2}{A_0^2}\left(\frac{1 + E/2mc^2}{1 + E/mc^2}\right)^2 \mathcal{H}_y(z,t) = 0. \quad (97)$$

If, in accordance with (17), we make the same (36) and (66), we obtain the solution:

$$v = \frac{Z^4 e^8 m + Z^2 e^4 A_0^2 \varepsilon^2 \left(\sqrt{m^2\left(64c^4 + \frac{Z^4 e^8}{A_0^4 \varepsilon^4}\right)} - 20mc^2\right)}{16 A_0^3 \varepsilon^2 \left(Z^2 e^4 - 6 A_0^2 c^2 \varepsilon^2\right)}$$

$$- \frac{8 A_0^4 c^2 \varepsilon^4 \left(\sqrt{m^2\left(64c^4 + \frac{Z^4 e^8}{A_0^4 \varepsilon^4}\right)} - 8mc^2\right)}{16 A_0^3 \varepsilon^2 \left(Z^2 e^4 - 6 A_0^2 c^2 \varepsilon^2\right)}. \quad (98)$$

$Z$ and $v$ are the only variables in equation (98), while all other sizes are constant, and $Z$ is the only independent variable. When $Z$ is relatively small, then instead of (98) applies the following expression:

$$v \approx \frac{m e^4}{8 A_0^3 \varepsilon^2} Z^2. \quad (99)$$

Energy, which is determined by (37), as well as the frequency, which is determined by (99), describes only one state of the atom, i.e., *reference state*, named in the article Physics Essays **16**, Ref. 2, 162 (2003) as the *state of reference*. All other existing states (*if such states, in general, exist*) of the atom should be described using only the remaining independent variable $Z$. This means that all other energies and frequencies of the atom should be described, for example, with $Z^*/s$ instead of $Z$, i.e., $Z \to Z^*/s$, where $s$ is any *positive real number* and $Z^*$ is the atomic number, such as was formerly $Z$. So from (37) and (99) we obtain:

$$eV = E_{em} \approx \frac{m e^4}{8 A_0^2 \varepsilon^2}\left(\frac{Z^*}{s}\right)^2, \quad (100)$$

$$v \approx \frac{m e^4}{8 A_0^3 \varepsilon^2}\left(\frac{Z^*}{s}\right)^2. \quad (101)$$

The difference of particular energy and frequency, characterized with different $s$, can be expressed as follows ($s' \geq s$):

$$E_{em(\Delta)} = E_{em(s)} - E_{em(s')} \approx \frac{m e^4 Z^{*2}}{8 A_0^2 \varepsilon^2}\left(\frac{1}{s^2} - \frac{1}{s'^2}\right), \quad (102)$$

$$v_{(\Delta)} = v_{(s)} - v_{(s')} \approx \frac{m e^4 Z^{*2}}{8 A_0^3 \varepsilon^2}\left(\frac{1}{s^2} - \frac{1}{s'^2}\right). \quad (103)$$

From here, with $c = \lambda v$, we obtain a formula similar to the Rydberg's formula:

$$\frac{1}{\lambda_{(\Delta)}} = \frac{1}{\lambda_{(s)}} - \frac{1}{\lambda_{(s')}}$$

(104)

$$\approx \frac{m e^4 Z^{*2}}{8 A_0^3 \varepsilon^2 c}\left(\frac{1}{s^2} - \frac{1}{s'^2}\right) = Z^{*2} R\left(\frac{1}{s^2} - \frac{1}{s'^2}\right).$$

Here $R$ is Rydberg's constant:

$$R = \frac{m e^4}{8 A_0^3 \varepsilon^2 c} = \frac{m e^4 c^3 \mu^2}{8 A_0^3}. \quad (105)$$

So, without using Bohr's postulates, I made expressions that are much more comprehensive than the expressions obtained by Bohr's model of the atom, because my model includes all the energy states of atoms ($s$ is a *positive real number*), while Bohr's model includes only certain quantum states $s = n$, ($n$ is only an *integer*). All, even the quantum states must be incorporated into the expressions (102), (103) and (104). Quantum states are special states of atoms in which there is a harmony of the mechanical motion of electrons and electromagnetic standing wave in an atom. Noting that the number $s$ in my model may be less than 1, it means that energy and frequency can be higher than those in the *ground state* $s = 1$. E.g., for the ground state of hydrogen atom, $s = 1$, while for neutrons and for some of the hyperons $s$ is significantly less than 1, (see Table III), [20]. For determining the state of particles ($s = Z^*/Z$) in Table III we use the solution, which derived from (37), ($eV_i = E_{em}$):

$$Z = \frac{2 A_0 \varepsilon}{e^2}\sqrt{\frac{E_{em}(E_{em} + 2mc^2)}{m(E_{em} + mc^2)}}, \quad (106)$$

and relativistic expression for the mass of an electron in motion $m/\sqrt{1-\beta^2} = m_x - m_p$, i.e.,

$$\beta = \sqrt{1 - [m/(m_x - m_p)]^2}, \quad (107)$$

where $m_x$ is the mass of the observed particles and $m_p$ is the mass of the proton. Atomic number for a single proton is $Z^* = 1$.

TABLE III. Determining the state of a particle $s = Z^*/Z$, on the basis of a change in mass of its electrons. For one proton $Z^* = 1$. Data are from Ref. 20 ($m = 0.511\,006$ MeV/$c^2$, $m_p = 938.256$ MeV/$c^2$).

| x | $m_x$ (MeV) | $\beta = v/c$ (1) | $r_a$ ($10^{-15}$) (m) | $-\mu_e$ ($10^{-26}$) (A m$^2$) | $K$ (keV) | $U$ (keV) | $E_{em}$ (keV) | $s = Z^*/Z$ (1) |
|---|---|---|---|---|---|---|---|---|
| H | 938.767 [a] | 0.007 295 6 | 52 941.319 | 927.649 | 0.013 6 | −0.027 | 0.013 6 | 1.000 234 |
| n | 939.550 [b] | 0.918 722 3 | 1.318 [c] | 2.908 [c] | 782.983 [d] | −1 092.187 | 309.203 4 | 0.007 364 |
| $\Lambda^0$ | 1 115.40 [b] | 0.999 995 8 | 0.008 | 0.0195 | 176 630.597 [d] | −177 140.122 | 509.525 0 | 0.005 965 |
| $\Xi^0$ | 1 314.30 [b] | 0.999 999 1 | 0.004 | 0.0092 | 375 527.898 [d] | −376 038.203 | 510.304 7 | 0.005 962 |

[a]The mass of hydrogen includes his ionization energy; $m_H = m_p + m + 13.6\,\text{eV}/c^2 = 938.767\,02$ MeV/$c^2$.
[b]Reference 20.
[c]Reference 21.
[d]Reference 22.

## V. CONCLUSION

Analyses of the Duane-Hunt's and Kulenkampf's measurement results confirm that the Duane-Hunt's law should be corrected. As a consequence Planck-Einstein's relation, de Broglie wavelength and Schrödinger's equation should also be corrected. This article proposes a solution for each of them, i.e., Duane-Hunt's law should be replaced with the expression (43), Planck-Einstein's relation with the expression (66), de Broglie's wavelength with equation (73), and, finally, Schrödinger's equation should be replaced with one of the corresponding equations (90) to (97). All these solutions were obtained using classical physics.

## VI. ACKNOWLEDGMENTS

The author would like to thank Dr. Vladimir Knapp for their continuous interest, Dr. Bogdan Zelenko for mathematical simplification of some expressions, Ms. Erica Vesic for editing this article in English, Mr. Nicolas Ranty for his French translation of the abstract, *Movens Ltd*, Zagreb and my family for general support. The Wolfram *Mathematica* software is used by courtesy of *Systemcom Ltd*, Zagreb, Croatia.

[1] P. W. Bridgman, National Academy of Science of the United States of America, Biographical Memoirs, Volume XVIII–Second Memoir, Biographical Memoir of William Duane, 1872-1935, Presented to the Academy at the Annual Meeting, 1936.
[2] M. Perkovac, Phys. Essays **15**, 41 (2002); **16**, 162 (2003).
[3] J. D. Jackson, *Classical Electrodynamics* (John Wiley & Sons, 2nd Edition, New York, 1975), p. 353.
[4] M. Perkovac, (Ref.2).
[5] Z. Haznadar and Z. Stih, *Elektromagnetizam* (Skolska knjiga, Zagreb, 1997), pp. 438, 552.
[6] R. Rüdenberg, *Elektrische Schaltvorgänge* (Verlag von Julius Springer, Berlin, 1923), p. 352.
[7] D. C. Giancoli, *Physics*, 2nd edition (Prentice Hall, Engelwood Cliffs, NJ, 1988), p. 700.
[8] T. Bosanac, *Teoretska elektrotehnika 1* (Tehnicka knjiga, Zagreb, 1973), p. 371.
[9] J. D. Jackson, *Classical Electrodynamics* (Ref. 3), p. 262.
[10] *Elektromagnetizam* (Ref. 5), pp. 202, 549.
[11] http://www.tpub.com/content/doe/h1019v1/css/h1019v1_52.htm
[12] Ch. Clapham, *The concise Oxford dictionary of mathematics* (Oxford University Press, Oxford New York, 996), p. 56.
[13] I. Sosic, *Primijenjena statistika* (Skolska knjiga, Zagreb, 2006), p. 415.
[14] W. Duane and F.L. Hunt, Phys. Rev. **6**, 166 (1915).
[15] H. Kulenkampff, Ann. d. Phys. **69**, 548, 548 (1922).
[16] *Primijenjena statistika* (Ref. 13), p. 389.
[17] E. W. Schpolski, *Atomphysik, Teil I* (VEB Deutscher Verlag der Wisenschaften, Berlin, 1979), p. 406.
[18] *Elektromagnetizam* (Ref. 5), p. 436.
[19] B. Wong, Phys. Essays **22**, 296 (2009).
[20] A. H. Rosenfeld, A. Barbaro-Galtieri, W. H. Barkas, P. L. Bastien, J. Kirz, and M. Roos, Rev. Mod. Phys. **36**, 977 (1964).
[21] S. Vlaicu, Phys. Essays **22**, 462 (2009).
[22] B. Yavorsky and A. Detlaf, *Handbuook of Physics* (Mir Publisher, 1975), p. 1051.